\documentclass[aps,showpacs,preprintnumbers,superscriptaddress,tightenlines]{revtex4}
\usepackage{amsmath}
\usepackage{dcolumn}
\newcommand{\nuc}[2]{\ensuremath{^{#2}}\textrm{#1}}
\newcommand{\micron}{\ensuremath{\mu\mathrm{m}}}

\begin{document}
\preprint{LA-UR-04-7476}
\title{Prompt Beta Spectroscopy as a Diagnostic for Mix in Ignited NIF Capsules}
\author{A.C. Hayes}
\affiliation{Theoretical Division}
\author{G. Jungman}
\affiliation{Theoretical Division}
\author{J.C. Solem}
\affiliation{Theoretical Division}
\author{P.A. Bradley}
\affiliation{Applied Physics Division}
\author{R.S. Rundberg}
\affiliation{Chemistry Division\\
Los Alamos National Laboratory, Los Alamos, NM 87545, USA}

\begin{abstract}
The National Ignition Facility (NIF) technology is designed to
drive deuterium-tritium (DT) internal confinement fusion (ICF)
targets to ignition using indirect radiation from laser beam
energy captured in a hohlraum. 
Hydrodynamical
instabilities at interfaces in the ICF capsule leading to mix between the DT fuel and
the ablator shell material
are of fundamental physical
interest and can affect the performance characteristics of the capsule.
In this Letter we describe new
radiochemical diagnostics for mix processes in ICF capsules with
plastic or Be (0.9\%Cu) ablator shells.
Reactions of high-energy tritons with shell material produce high-energy 
$\beta$-emitters.
 We show that mix between the
DT fuel and the shell material enhances high-energy prompt beta
emission from these reactions by more than an order of magnitude over
that expected in the absence of mix.
\end{abstract}

\pacs{52.57.Fg,25.10.+s, 25.55.-e}

 \maketitle

\section{INTRODUCTION}
The projected fusion yields at NIF suggest that radiochemical
analysis of neutron and charged-particle {\it incidental}
reactions could provide important diagnostics for physical
processes in the implosion and burn of an ICF capsule. In
particular, such reactions can probe mixing at the capsule
interfaces owing to hydrodynamic instabilities. Materials in
current capsule designs can be used as effective radiochemical
detectors for mixing processes without introducing additional
dopants. For example, Stoyer {\it et. al.} pointed out that the
$^8$O($\alpha$,n)$^{21}$Ne reaction can be a sensitive probe of
plastic-shell mixing \cite{stoyer}.  In the present paper we
concentrate on capsule designs with either plastic or beryllium
shells. The capsules involve a layer of DT ice
approximately \(80\ \micron\) thick with \(\rho=0.25\ \mathrm{g \
cm^{-3}}\),
an interior gas fill \(\rho=0.005\ \mathrm{g\ cm^{-3}}\),
and an ablator layer thickness of approximately \(155\ \micron\).
With a peak hohlraum temperature of 330 eV, the capsules are
expected to reach a peak \(\langle \rho R \rangle_{DT}\sim 1.2\
\rm{ g\ cm^{-2}}\) and achieve ignition with a  yield up to about
18 MJ.

A microscopic understanding of mix remains one of the key issues
for ICF \cite{Radha}. We propose to study mix of the ablator shell
into the DT fuel by measuring the number of interactions of
tritons with the shell material. The promising reactions emit
high-energy $\beta$s, and we propose prompt $\beta$ spectroscopy as a
new probe of mix. We present the physics underlying the
sensitivity of these reaction rates to mix and  estimates for the
reaction yields under different mix scenarios.

\section{DETECTION}
The set of core NIF diagnostics is expected to include a Debris
Collector and Rad-Chem Station \cite{miller}. A suite of
collectors will be placed on the target chamber wall, covering
multiple lines of sight. Variations of this capability could allow
for prompt diagnostics.
%Assuming realistic geometries, we estimate that a fraction of
%about $10^{-6}$ of the nuclides produced will be collected.
%This corresponds to a \(1\ \mathrm{cm}^2\) detector area at a
%distance of 2.8 m; of course, this collection efficiency can
%vary by as much as a factor of four, depending on the actual
%location of the collector.
An alternate scheme that would allow a higher collection fraction
is a variation of the gas collection system designed by Lawrence
Livermore National Laboratory \cite{stoyer} and installed at
OMEGA. The reactions of interest in this letter would require the
gas collection system to be fast enough to isolate samples with
half-lives on the order of seconds.

The triton induced reactions of most interest for our
proposed studies of mix produce high-energy $\beta$ emitters.
For the plastic ablator shells we consider the
$^{18}$O$(t,n)^{20}$F($\beta^-$),
and
$^{13}$C$(t,\alpha)^{12}$B($\beta^-)$
reactions, and for the Be ablator shells the
$^9$Be$(t,\alpha)^8$Li($\beta^-$) and
$^9$Be($t,p)^{11}$Be($\beta^-$) reactions. 

The  energy of the $\beta$s is high enough for them to be clearly
distinguishable from prompt laser ejected electrons. Table III
summarizes the properties of all the reactions of interest. In the
cases of $^{20}$F and $^{11}$Be, the additional
coincidence $\gamma$-ray aids detection and distinction from
$\beta$s of similar energy. The cross sections are not known for
all the reactions. 
The
\nuc{Be}{9}(t,$\alpha$)\nuc{Li}{8}($\beta^-$)  reaction has a
cross section of about 17 mb at 0.5-MeV  and about 185 mb at 1.5
MeV \cite{nam}. The \nuc{Be}{9}(t,p)\nuc{Be}{11}($\beta^-$)
reaction has a threshold at 1.55 MeV and a cross section of about
1 mb at 3 MeV. The $^{18}$O(t,n)$^{20}$F cross section is not known but is expected
to be reasonably large \cite{becker}.

A thick aluminum window on the beta spectrometer in the Debris
Collector would act as a high-energy gate to eliminate the
background, and the different life-times of the $\beta$ decays of
interest would be used to distinguish the reactions.
The number of decays
needed to provide a half-life signature and counting statistics
better than 10\% is about 10$^4$ atoms. The solid angle for debris
collection is about 10$^{-6}$, assuming a 1 cm$^2$ debris
collector. The beta counter efficiency on the order of 10\%
includes the effect of the aluminum absorber. The minimum number of atoms
needed to provide the requisite number of decays is thus 10$^{10}$ atoms
for a debris collection system.
This limit could be lowered a few orders of magnitude by
increasing the effective area for debris collection.

The \nuc{F}{20} gaseous products could be retrieved by
a gas collection system. The efficiency for fluorine collection
should be high because the NIF target chamber is constructed of
aluminum.  The estimated time for 99\% collection of gases is 90
seconds.  The reduction in the number of atoms during collection
is  3.5 x 10$^{-4}$ for \nuc{F}{20}.
The efficiency for an internal (4$\pi$) gas proportional counter
is near 100\%. Therefore, the minimum number of atoms
for a detection is about 
 10$^8$ for \nuc{F}{20}.

\section{MIX REACTION RATE ENHANCEMENT }

\subsection{Knock-on Triton Fluence}
The energetic tritons in the above reactions are produced by
collisions with high energy neutrons from the central DT burn. We
call these {\it knock-on} tritons. The number of knock-on tritons
scales with the high-energy neutron production. Some of these
energetic tritons go on to fuse with deuterium, some will escape,
but a significant fraction will react with detector nuclei in the
shell. 
Stopping of these energetic tritons reduces the probability of 
reaction with nuclei in the shell.
If the shell material is intimately mixed with the fuel, the
reaction yields are enhanced by the ratio of the mixing length (the thickness of the 
mixed region) to the triton 
stopping length in the unmixed fuel. Therefore, the total number of reactions of knock-on tritons
with shell material is a measure of the mixing.

In the typical implosion scenario, ignition occurs near the center
of the DT volume, in the {\it hotspot} region. The DT combustion
progresses outward from the hotspot, finally reaching the shell
region and completing. During most of the burn, the interfacial
region near the shell is inert, evolving hydrodynamically under
the external driving force and the internal pressure. During the
interior burn, the interfacial region is bathed in 14 MeV neutrons
which produce a spectrum of energetic knock-on tritons. Because
the triton stopping length in the DT plasma is relatively short,
only knock-on tritons produced near the interfacial region
contribute to the high-energy part of the triton spectrum as seen
by the shell material. A radially localized source of high-energy
tritons is created near the shell, absent of complications due to
the DT-burn wave, which is still near the center of the capsule.

For the nominal design discussed in this paper, approximately
\(10^{19}\) high-energy (14 MeV) neutrons are produced by DT burn.
The shape of the neutron spectrum does not deviate significantly
from a thermally-broadened 14-MeV peak because the system is quite
thin to high-energy neutrons. The differential production rate for
knock-on tritons at a point in the fuel is $\phi_n n_t
(d\sigma_{nt}/dE)$, where \(\phi_n\) is the neutron flux, \(n_t\)
is the triton density at the specified point, and
\(d\sigma_{nt}/dE\) is the (normalized) distribution of tritons in
the final state due to collisions with 14-MeV neutrons, a known
kinematic function. The effective size (radius) of the triton source and
the distortion of the triton spectrum depend on the Coulomb drag.
Using an analytic approximation for path length \cite{Longmire},
integrating over the position of triton production, and assuming
that the triton density and the electron temperature in the fuel
are approximately constant over a triton trajectory, we obtain the
differential knock-on triton fluence
\begin{equation}
   \frac{d\psi_t}{dE} = \frac{\theta_e^{\frac{3}{2}} \psi_n n_t}{c_R n_e
\sqrt{E}}\ \sigma_{nt}(\ge E), \label{eqn:tritonfluence}
\end{equation}
where \(\theta_e\) is the electron temperature, \(\psi_n\) is the
neutron fluence, \(n_e\) is the electron density, \(c_R \simeq
1.5\times 10^{-20} \mathrm{keV}^2 \mathrm{cm}^2\) and
\(\sigma_{nt}(\ge E) = \int_E^{E_{\mathrm{max}}} dE\,
d\sigma_{nt}/dE\) is the integral of the microscopic knock-on
triton distribution over triton energies greater than the given
energy \(E\) ($E_{\mathrm{max}}\simeq 10.5$ Mev).

\subsection{Scenario I: No-Mix, Smooth and Wrinkled Shells}
First we consider the interaction of energetic tritons with a
thick unmixed region of shell material. The tritons cross the
surface of the material with a differential fluence given by
Eq.(\ref{eqn:tritonfluence}). The rate of triton reactions with the
shell nuclei is small compared to the rate of energy loss, so that
the probability of reaction of a triton of energy \(E\) with a
shell nucleus is
\[
   P(E) \simeq \hat n \int_0^E dE'
\frac{\sigma_{\mathrm{t,shell}}(E')}{-(dE/dx)_{\mathrm{shell}}}.
\]
Here $\hat n$ is the number density of shell ions that could serve
as radio-chemical detectors ($^9$Be, $^{13}$C, or $^{18}$O) and
\(\sigma_{\mathrm{t,shell}}\) is the energy-dependent
cross-section for the triton reaction with the shell nucleus of
interest. Assuming that the shell electron density and
temperature, \(\hat{n}_e\) and \(\hat{\theta}_e\), are constant
over the the distance of a triton stopping length, we then have
\[
   P(E) \simeq 2 \frac{{\hat n}}{\hat{n}_e} \frac{\hat{\theta}_e^{\frac{3}{2}}}{c_R}
     \int_0^E dE' \frac{\sigma_{\mathrm{t,shell}}(E')}{\sqrt{E'}}.
\]
If we adopt the simplest physical model for a cross section with
threshold as \(\sigma(E) = \sigma_0 \Theta(E \ge E_*)\), where
\(\Theta\) is the Heaviside step function, the probability for a
$t+shell$ reaction becomes,
\begin{equation}
   P(E) \simeq 2 \frac{\hat{n}}{\hat{n}_e} \frac{\hat{\theta}_e^{\frac{3}{2}}}{c_R}
     \sigma_0 \left(\sqrt{E} - \sqrt{E_*}\right) \Theta(E \ge E_*).
\label{eqn:PE}
\end{equation}
The total number of reactions of high-energy tritons with shell
nuclei is obtained by integrating the product of
Eq.(\ref{eqn:tritonfluence}) and Eq.(\ref{eqn:PE}), and
multiplying by the shell area. We also need to introduce a
geometric flux factor $g(E)$ which is dependent on the triton
energy, since higher energy tritons are produced preferentially
in the forward (radial) direction. For a surface described by
periodic wrinkling with maximum normal deflection angle $\chi$,
the geometric flux factor is given by\\
\[
g(w,\chi) =
\begin{cases}
   \cos\eta, \,\, 0 \le |\eta| \le \pi/2 - \chi &  \\
   \left[(\chi - |\eta| + \pi/2)\cos\eta
   + \ln\left({\sin |\eta|}/{\cos\chi}\right) {\sin|\eta|}\right]/(2\chi),
& \text{otherwise},
\end{cases}
\]
where \(w = E/E_{\mathrm{max}}\) and $\cos\eta=\sqrt{w}$. The
number of reactions with the shell material is then
\[
   \mathcal{R} \simeq  \hat{f}{f}\sigma_0 N_n
     \frac{(\theta_e \hat{\theta}_e)^{\frac{3}{2}}}{c_R^2}
     \int_{\ge E_*} dE\, \sigma_{nt}(\ge E) g(E)
       \left(1 - \sqrt{\frac{E_*}{E}}\right),
\]
where $\hat{f}={\hat{n}/\hat{n}_e}$ is the ratio of the density of
\emph{detector} nuclei to the density of free electrons in the
shell, ${f}={{n}_{t}/{n_e}}\simeq 0.5$ is the ratio of density
tritons to density of free electrons in the fuel, and \(N_n\) is
the total number of 14-MeV neutrons produced ($\psi_n\times$
area). The distribution of knock-on tritons is expressed as the
product of a total cross section $(\sigma^{\rm{TOT}}_{nt})$ and a
triton spectral shape, which can be approximated as
\[
   h_{nt}(w) = \frac{\sigma_{nt}(\ge w)}{\sigma_{nt}^{\mathrm{TOT}}}
        \simeq 1 - \frac{1}{\pi}
                   \left[ \sqrt{w(1-w)} + 2 \sin^{-1} \sqrt{w}\right].
\]
The number of detector activations is then
\[
\mathcal{R} \simeq
    4.6\times 10^{13}
    \hat{f} {f} \frac{N_n}{10^{19}}
    \frac{\sigma_0}{100\ \mathrm{mb}}
    \frac{\sigma_{nt}^{\mathrm{TOT}}}{1\ \mathrm{b}}
    \left( \frac{\theta_e \hat{\theta}_e}{1\ \mathrm{keV}^2}
\right)^{\frac{3}{2}}
    I(w_*,\chi).
\]
where
\[
   I(w_*,\chi) = \int_{\ge w_*} dw\, h_{nt}(w) \, g(w,\chi)\,
                      \left(1 - \sqrt{\frac{w_*}{w}}\right),
\]

The sensitivity of the  $t+shell$ reaction rate to the threshold
energy and the shape of the fuel/shell interface is carried in the
integral $I(w_*,\chi)$. We estimate that in the shell region the
detector nucleus to electron ratio is
  $\hat{f}\simeq 0.25, 2\times 10^{-3}, 5\times 10^{-5}$ for
\nuc{Be}{9}, \nuc{C}{13} and \nuc{O}{18}, respectively. As shown in Table I,
we find less than a factor of two increase in the predicted
$t+shell$ reaction rates for a highly rippled interface over that
for a smooth interface.

\begin{table}
\begin{tabular}{c|ccc}
\multicolumn{4}{c}{\textbf{Non-Dimensional Integral $I(w_*,\chi)$}} \\
\hline
     Threshold       &  \multicolumn{3}{c}{$2\chi/\pi$}  \\
 \(w_*=E_*/E_{max}\) &        0.0 & 0.5 & 0.9            \\
\hline
 0.75  & 0.002 & 0.002 & 0.002 \\
 0.50  & 0.011 & 0.011 & 0.015 \\
 0.30  & 0.033 & 0.033 & 0.047 \\
 0.25  & 0.042 & 0.043 & 0.061 \\
 0.20  & 0.054 & 0.044 & 0.078 \\
 0.15  & 0.069 & 0.070 & 0.103 \\
 0.10  & 0.088 & 0.091 & 0.136 \\
 0.05  & 0.118 & 0.122 & 0.189 \\
 0.01  & 0.161 & 0.172 & 0.276 \\
 0.0   & 0.197 & 0.217 & 0.361 \\
\hline
\end{tabular}
\caption{Values for the non-dimensional integral $I(w_*,\chi)$
appearing in the reaction rate for a  no-mix DT fuel/shell
interface. As the shape of the interface ($2\chi/\pi)$ goes from
flat ($2\chi/\pi$=0) to highly wrinkled ($2\chi/\pi$ =0.9) there
is less than a factor of two increase in the reaction rate.}
\label{tab:Ia}
\end{table}

\subsection{ Scenario II: Chunk Mix}
Next we consider the case of so-called {\it chunk} mix, where
micron-scale pieces of shell material are injected into the fuel
region by shell instabilities. This case is relatively easy to
treat as a modification of the unmixed case treated above.
Specifically, we assume that the chunks are large enough to stop
energetic tritons within their volume and that the spacing between
chunks is larger than the triton stopping length in the fuel. In
this situation there is no shielding effect for triton production
and the total number of reactions is enhanced by a simple
geometric factor (\(\mathcal{R} =\mathcal{E}_{chunk}\mathcal{R}\))
reflecting the increased surface area exposed to the triton flux.
We find an enhancement factor,
\[
   \mathcal{E}_{chunk} \simeq
      1 + \frac{3}{4\pi} \frac{\ell}{R_c} \frac{\rho_c}{\rho_{\text{shell}}},
\]
where \(\ell\) is a mixing length, giving the size of the region
over which chunks are dispersed, \(R_c\) is the average size of
chunks, and \(\rho_c/\rho_{\text{shell}}\) gives the fraction of
shell material mixed into the fuel as chunks, as a fraction of the
nominal shell material density. This final factor can be quite
small in physically reasonable situations, although it is
partially compensated by the ratio \(\ell/R_c\), which can be
expected to be large.

Hydrodynamic simulations typically show interface instabilities
leading to the formation of highly non-linear interface
structures, not unlike our highly-wrinkled no-mix scenario.
Assuming that these structures can  break-off and lead to chunk
mix, the chunk radii would be of the order of a micron. With an
assumed mixing  length $\ell\sim 4 \micron$m, this scenario would
imply \(\ell/R_c\sim 1\). Thus, no enhancement over the
highly-wrinkled-but-unmixed interface would be expected for {\it
large} chunk-mix. On the other hand, if the chunks mixed into the
gas were small (order of 0.1 \micron)  the enhancement factor
\(\mathcal{E}_{chunk}\) could be as large as a factor of about
two.

\subsection{Scenario III: Atomic Mix}
Finally we consider the situation in which an atomically mixed
region exists near the shell interface, characterized by a mixing
length \(\ell\). In this case the high-energy tritons interact
directly with shell ions mixed into the fuel. The total number of
reactions in the mix region is given by the expression,
\[
   \mathcal{R}_{\rm mix} = \int_{\mathrm{mixed}\atop{\mathrm{region}}} dr \int dE\,
       \frac{d\psi_t}{dE}\,
       \sigma_{\mathrm{t,detector}}(E)\,
       \mathrm{Area}(r)\,
       \hat{n}(r),
\]
where \(d\psi_t/dE\) is again the differential triton fluence,
given by Eq.(\ref{eqn:tritonfluence}). For the purpose of our
estimate it is reasonable to assume the basic temperature and
density parameters are constant within the mix region. In this
case we obtain
\[
   \mathcal{R}_{\rm{mix}} \simeq
     2.7\times 10^{14} \hat{f}
     \frac{N_n}{10^{19}}
     \frac{n_t}{10^{26}\ \mathrm{cm}^{-3}}
     \frac{\ell}{4\ \micron}
     \frac{\sigma_0}{100\ \mathrm{mb}}
     \frac{\sigma_{nt}^{\mathrm{TOT}}}{1\ \mathrm{b}}
     \left(\frac{\theta_e}{1\ \mathrm{keV}}\right)^{\frac{3}{2}}
     I_{\rm{mix}}(w_*),
\]
where
\[
   I_{\rm{mix}}(w_*) = \int_{\ge w_*} \frac{dw}{\sqrt{w}} h_{nt}(w).
\]

In a realistic physical picture, the density of the DT gas and the shell ions are more
likely to vary (as opposed to remaining constant) across the mix
region. For a linear mixing profile, this variation reduces the
$t+shell$ reaction rate by a factor of 1/6, which is included
in above estimate.

$I_{\rm{mix}}(w_*)$ is easily evaluated and some results are given
in Table \ref{tab:Im}. The values for this integral are an order
of magnitude larger than the values obtained for $I(w_*,\chi)$.
This enhancement reflects the immersion of the shell material
in the fuel, which gives a volume enhancement proportional to
the ratio of the mixing length to the triton stopping length.

\begin{table}
\begin{tabular}{cc}
   \(w_* = E_*/E_{max}\) & \(\quad I_{\mathrm{mix}}(w_*)\)  \\
\hline
   0.50      &   0.12 \\
   0.40      &   0.18 \\
   0.30      &   0.25 \\
   0.20      &   0.36 \\
   0.15      &   0.43 \\
   0.10      &   0.52 \\
   0.05      &   0.66 \\
   0.01      &   0.87 \\
   0.00      &   1.06 \\
\hline
\end{tabular}
\caption{Values for the non-dimensional integral \(I_{\mathrm{mix}}(w_*)\)
appearing in the reaction rate for atomically mixed interfaces.
The reaction rates are about an order of magnitude larger
than the unmixed case.}
\label{tab:Im}
\end{table}

\section{CONCLUSIONS}
Charged particle reactions leading to prompt high-energy
$\beta$-decay signals can provide an important diagnostic for
mixing in ICF capsules. The predicted number of such reactions, at
the yields and neutron fluences expected at the NIF facility,  is
sufficient for a detectable radiochemical signature of mixing
processes. Our estimates for the interaction of high-energy
knock-on tritons with  shell material suggest that an adequate
number of beta-emitters are produced to provide a unique prompt
diagnostic for mix between the fuel region and the shell material.
The reaction yields for the $t+shell$ reactions of interest are
enhanced by the ratio of the mixing length to the triton stopping length, and
this leads to increased reaction yields of more than an order of 
magnitude over the no-mix situation.
Table III lists and summarizes our estimated $t + \nuc{Be}{9}$,
$t + \nuc{O}{18}$, and $t + \nuc{C}{13}$ reaction yields for the
different mix scenarios.

\newcolumntype{d}[1]{D{.}{.}{#1}}
\begin{table*}
\begin{tabular}{c|r|rcr|l|c|lll}
\multicolumn{10}{c} {\bf Expected Reaction Yields}\\
\hline
Reaction & Q-value & \multicolumn{3}{c|}{\(\beta\) energy} & Half-life & \(\sigma_0\) & No Mix & Chunk Mix & Atomic Mix \\
\hline
& & & & & & & & &\\
\nuc{Be}{9}\((t,\alpha)\)\nuc{Li}{8} & 2.93 MeV  &  13.0 MeV & &        & 840 ms & 200 mb & \(1.4\times10^{12}\) & \(2.8\times10^{12}\) & \(9.3\times10^{13}\) \\
\nuc{Be}{9}\((t,p)\)\nuc{Be}{11}     &-1.17 MeV  & 11.5 MeV & & 54.7\% & 13.8 s & 1 mb   & \(4.0\times10^9\)    & \(8.2\times10^{9}\)  & \(3.6\times10^{11}\) \\
                                   & &  9.4 MeV  & (+2.13 MeV $\gamma$) & 31.4\% & & & & \\
\nuc{O}{18}\((t,n)\)\nuc{F}{20}      &  6.10 MeV & 5.3 MeV & (+1.63 MeV $\gamma$) & & 11.0 s & 100 mb & \(1.4\times10^8\) & \(2.8\times10^8\) & \(9.3\times 10^{9}\) \\
\nuc{C}{13}\((t,\alpha)\)\nuc{B}{12} & 2.28 MeV  &13.37 MeV & & & 20 ms & 100 mb & \(5.4\times 10^{9}\) & \(1.1\times 10^{10}\) & \(3.5\times 10^{11}\) \\
\hline
\end{tabular}
\label{tab:Ir} \caption{Estimated number of reactions for the
different mix scenarios. For the first two reactions we took the
cross sections to be 200 mb and 1 mb, respectively.
Little information is available for the \nuc{O}{18} and \nuc{C}{13}
reactions, and they are arbitrarly set to 100 mb. We took $\sigma_{nt} = $ 1b, 
 $\theta_e$= 1 keV, and $\hat{\theta}_e $= 1 keV. For the positive-Q reactions we took
$w_*$ = 0.05 to account for the Coulomb barrier. To estimate the yields for the chuck mix scenario
we took $2\chi/\pi =0.5 $ and $\mathcal{E}_{chunk} = 2.0 $. }
\end{table*}

\end{document}